\documentclass[11pt,a4paper]{article}

\usepackage[T1]{fontenc}
\usepackage[utf8]{inputenc}
\usepackage{lmodern}
\usepackage[margin=1in]{geometry}
\usepackage{microtype}

\usepackage{amsmath,amssymb,amsthm,mathtools}
\usepackage{graphicx}
\usepackage{booktabs}
\usepackage{siunitx}
\usepackage{xcolor}

\usepackage{authblk}

\usepackage{pdflscape}

\usepackage{float}

\usepackage[square,numbers,sort&compress]{natbib}
\usepackage[hidelinks]{hyperref}
\usepackage[nameinlink,capitalize,noabbrev]{cleveref}

\usepackage{tabularx}
\usepackage{threeparttable}
\usepackage{booktabs}

\usepackage[newfloat,frozencache,cachedir=.]{minted}
\setminted{fontsize=\small,breaklines,autogobble,tabsize=2}
\SetupFloatingEnvironment{listing}{name=Code Listing}

\title{Phantom types for robust hierarchical models with \texttt{typegeist}}

\author[]{Daniel O'Hanlon}
\affil[]{Imperial College London}
\affil[ ]{\texttt{dohanlon@imperial.ac.uk}}

\date{\vspace{-2em}}

\begin{document}
\maketitle

\begin{abstract}
  In Bayesian hierarchical models, group-level parameter arrays must be mapped to the observation axis, often using explicit indexing. In complex models with numerous incompatible data and parameter sets, this introduces the potential for bugs, as indexing with the incorrect indices typically fails silently. Here we present \texttt{typegeist}, a type system for Python that uses static type analysis to enable specification and enforcement of data-parameter-index correspondences. We show how this can be used with common probabilistic programming frameworks to help guarantee model correctness with minimal run-time overhead.
\end{abstract}

\section{Introduction}

A key driver in the design of modern programming languages is to make it as easy as possible to write bug free software, and type systems enable this by embedding provable mathematical guarantees about code behaviour. However, bugs that arise in statistics and data analysis software are far from those covered by the concerns of traditional programming language design. Programming languages are more often designed to address problems arising from memory safety~\cite{matsakis2014rust}, concurrency~\cite{cox2022go}, or side effects~\cite{hudak1992report}, than the validation of inputs or internal logic.

In machine learning in particular, unwieldy multidimensional tensors invite subtle mistakes due to the existence of multiple conflicting conventions on the meaning of each dimension, along with the ability to reorder and reshape with total freedom. To reduce the burden on the user to keep track of the semantics of these many dimensions, `named tensors' have been proposed to enforce consistency across operations~\cite{chiang2021named,hoyer2017xarray}. For data, tools that perform run-time schema, data type, and value domain validation have also been gaining popularity~\cite{niels_bantilan-proc-scipy-2020,pydantic_docs_2025}.

In hierarchical models, information from different levels of a problem can be individually described, potentially with their own observations, and feed into the probability distributions of a more complex model at a higher level of abstraction. Particularly in fields where mechanistic models and high-throughput data acquisition techniques exist, such as molecular biology ~\cite{lopez2018deep} and astrophysics~\cite{mandel2022hierarchical}, these are seeing increasing popularity, especially in conjunction with Bayesian deep learning.

\begin{figure}
\label{fig:intro}
\centering
\includegraphics[width=0.99\textwidth]{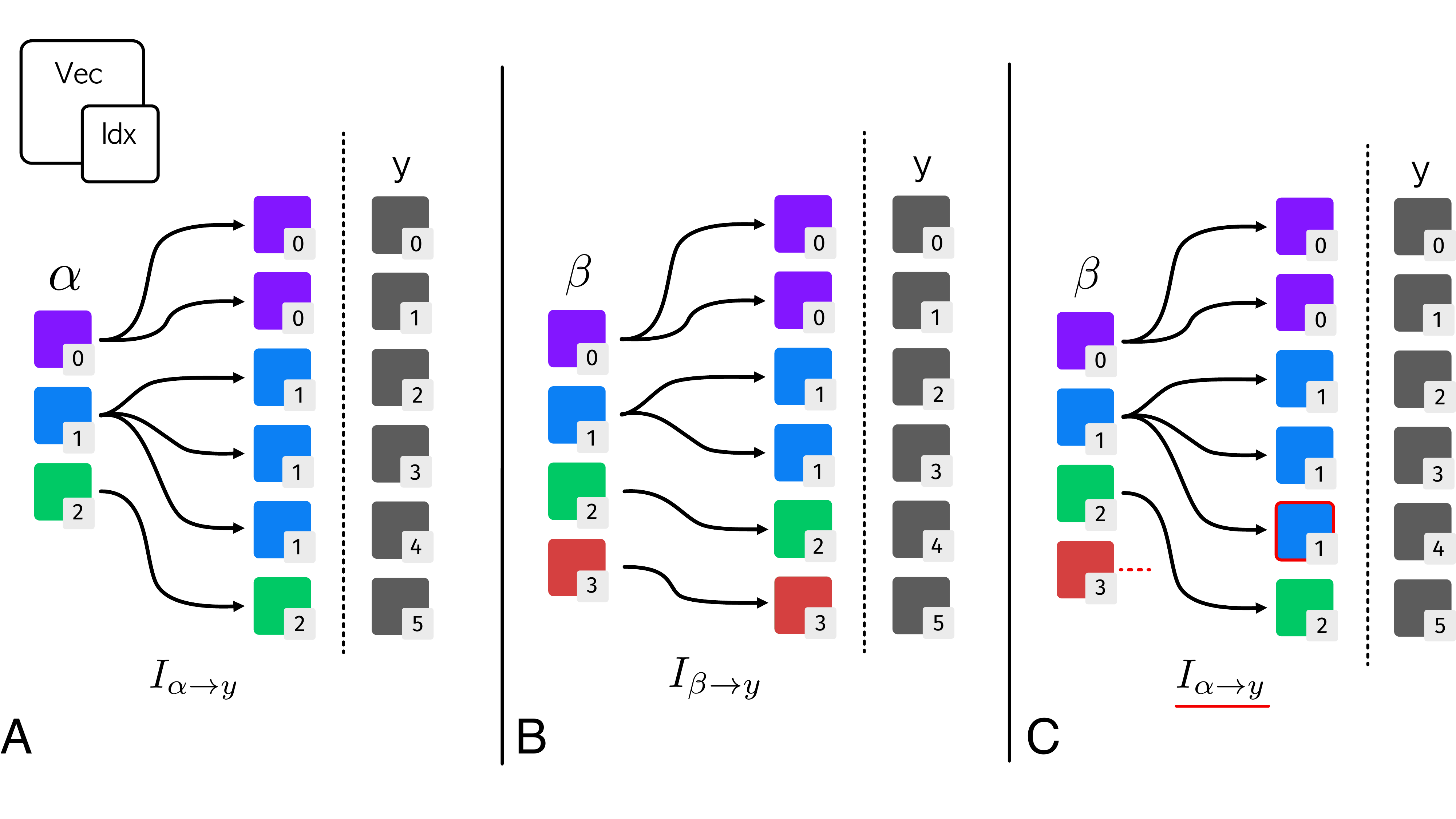}
\caption{A model with observations, $y$, and covariates $\alpha$ and $\beta$, where the colours indicate categories of covariates that have common distributional parameters. In the simplest case, $\alpha$ and $\beta$ both inform the observations, but have different mappings, $I_{\alpha \rightarrow y}$ and $I_{\beta \rightarrow y}$. In \textbf{A} and \textbf{B}, these parameters are broadcast to the observation axis via a \texttt{gather} operation using the correct corresponding indices. However, if incorrect indices are mistakenly used, for example in \textbf{C} where the $\beta$ covariates are mapped using the $\alpha$ indices, the corresponding contribution to the likelihood is nonsensical, where some variables do not appear and others appear where they should not.}
\end{figure}

In these models, a parameter, $\alpha$, describes a population-level covariate which must be `broadcast' to a likelihood that matches the full shape of the observation vector, $y$, of length $|y|$. A set of indices $I_{\alpha \rightarrow y}$ therefore maps the domain of the population parameter to the domain of observations, $y$, $y \leftarrow \alpha[I_{\alpha \rightarrow y}]$. In a simple model with only one set of indices, this is unlikely to cause issues. However, when introducing an additional parameter, $\beta$ with indices $I_\beta$, and describing a different aspect of the observations, we introduce the potential for errors. For example, transposing index arrays of the same length by using $I_\beta$ to index $\alpha$ into $y$, $y \leftarrow \alpha[I_{\beta \rightarrow y}]$, can fail silently (providing $I_{\beta}$ doesn't index off the end of the $\alpha$ array), yet provides nonsense observation parameters (Figure~\ref{fig:intro}). These two issues, amongst others that will be presented later, can be avoided by operations which are required to respect semantics, in addition to the structure, of the parameter vectors and the index-defined transformations between them.

One particular avenue for improving correctness in statistical models specifically are phantom types~\cite{phantom-types}, which go beyond the typical notion of a type, and contain values that can be tested by predicates \emph{statically}: without having to execute the code. For models that implement complex data semantics, such as hierarchical models or those with complex experimental designs, these can ensure that when likelihood functions are constructed, each contributing level of the hierarchy or design contributes in the correct manner. These can be thought of as a kind of contract described by the user, indicating which transformations will be allowed on which objects. By creating a directed graph of relations, we restrict transformations only to those which have meaningful results defined by these relations. These relations are not limited only to those that describe measurement data however, as powerful constraints on the model parameters can also be obtained using the form of experimental designs or other structural relationships.

Here we show that by introducing types of \texttt{Vec}, \texttt{Idx}, and \texttt{Map}, which can be specialised by annotations of the (parameter) vector semantics and allowed transformations between them, we eliminate many pitfalls in the construction of hierarchical models. This improves code clarity, eases debugging, and increases confidence in model correctness. The formalism described here is implemented in the \texttt{typegeist} Python package~\cite{typegeist}, utilising the \texttt{mypy} static type checker~\cite{mypy}. This is compatible with most popular probabilistic programming frameworks such as \texttt{Pyro}~\cite{bingham2019pyro}, \texttt{NumPyro}~\cite{phan2019composable}, \texttt{PyMC}~\cite{abril2023pymc}, and \texttt{TensorFlow Probability}~\cite{dillon2017tensorflow}, however for brevity all examples here will be using \texttt{Pyro} (where features specific to \texttt{Pyro} will be indicated).

\section{Hierarchical models}
\label{sec:hm}

Hierarchical models incorporate group-level effects within the generative process. Often this is enabled by being able to insert additional information into the model, either by having additional data at a particular level of the hierarchy, or specifying how these latent components generate the observed data. The additional specification is a form of regularisation, and as such these models are generally more robust and interpretable.

One of the simplest such models is given by Gelman~\cite{gelman2006multilevel}, and is used to model residential radon levels in the USA. This model assumes that radon levels are increased by having a basement in the home, and the county-wide average of uranium in the soil (radon gas is produced by uranium decay). For property $i \in \{1, ..., N\}$ in county $j \in \{1, ..., J\}$, the measured radon level $y_{ij}$ is modelled in terms of a county-level effect for county $j$, $\alpha_j$, plus term $\beta$ depending on whether the property has a basement, with $x_{ij}$ as an indicator. The county-level contribution is described by a constant offset, $\gamma_0$, plus a variable $\gamma_1$ that controls the effect of the logarithm of the uranium level in county $j$, $u_j$. The likelihood is constructed by assuming these are normally distributed,
\begin{align}
y_{ij} &\sim N(\alpha_j + \beta x_{ij}, \sigma_y^2) \\
\alpha_j &\sim N(\gamma_0 + \gamma_1u_j, \sigma_\alpha^2),
\end{align}
with a prior on variance the county-level term of $\sigma_\alpha^2$, and radon measurement variance $\sigma_y^2$. Priors on the random variables of interest, $\beta$, $\gamma_0$, and $\gamma_1$, are uniform distributions with reasonable bounds. As Gelman notes, this model achieves more favourable results compared to full-pooling (single $\alpha$ for all counties) and no pooling (independent $\alpha_j$ for all counties), by allowing the degree of the constraint to vary naturally with the data.

\begin{figure}
\label{fig:radon_model_data}
\centering
\includegraphics[width=0.75\textwidth]{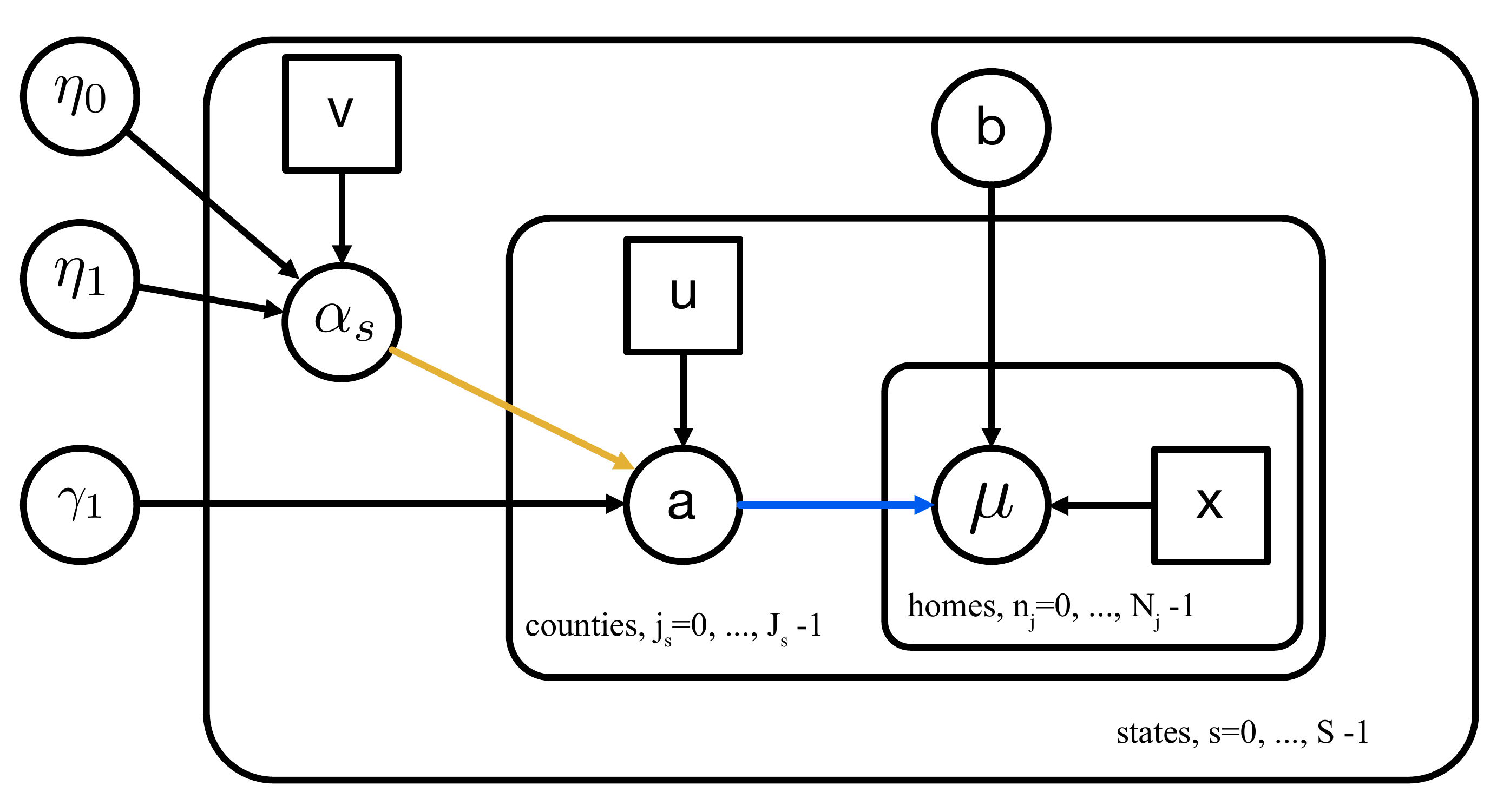}\\
\vspace{1cm}
\includegraphics[width=0.90\textwidth]{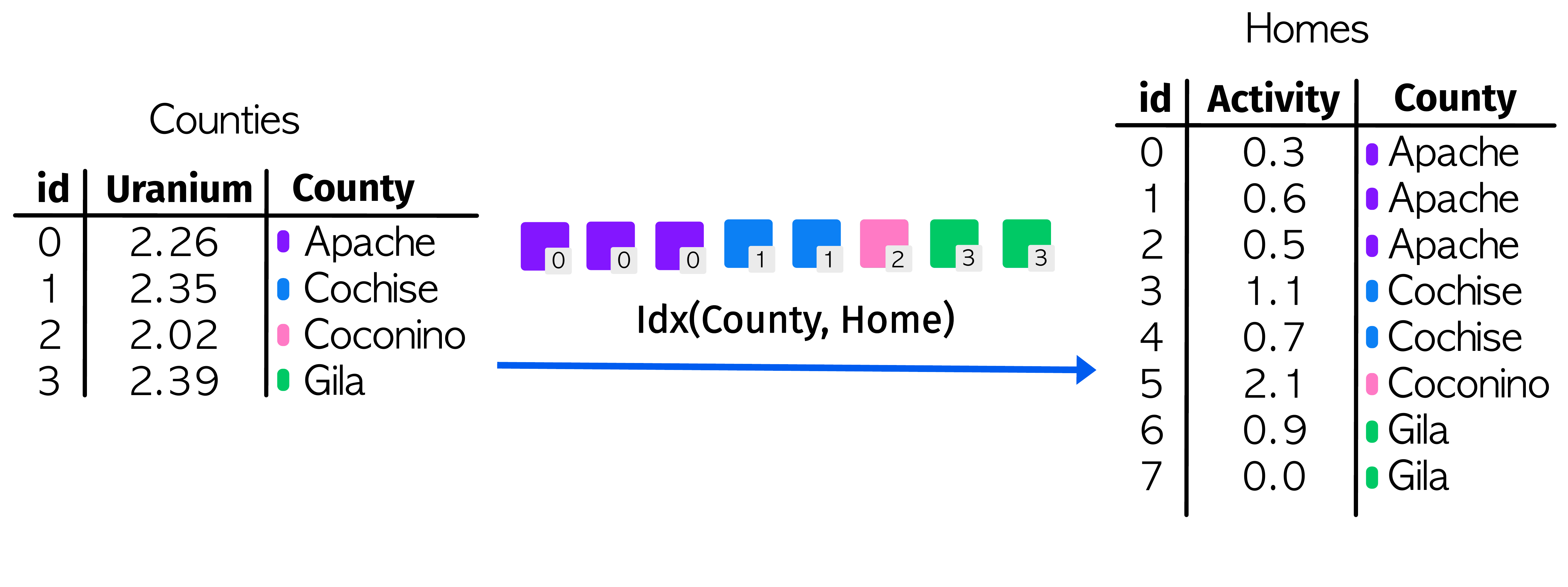}
\caption{A plate diagram for the three-level Gelman radon model (\textbf{A}). Circles represent random variables, squares represent constants, the large rectangular `plates' encompass entities with a common index, and arrows indicate parametric dependencies. Here the arrow between the county-level effect and the home-level effect is indicated in blue. As it traverses the plates, a mapping between the county-level indices and the home-level indices must be constructed (\textbf{B}). The additional mapping from states to counties is given in orange. Indices are shown as mappings that transform between the county to home space, as in the plate diagram.}
\end{figure}

This hierarchical model has two levels, one for how the measured uranium levels affect the county-level contribution, $\alpha_j$, and one for how the county-level contribution and the basement covariate affect the measured radon level, $y_{ij}$. In general, $J \leq N$ as there is at least one measurement per county, but there could be many more, and as such this is not a one-to-one relationship. A schematic of the dependencies in this model can be seen in Figure~\ref{fig:radon_model_data}. 

Whilst models \emph{can} be constructed with arbitrary logic that independently relates an individual observation with its corresponding parameters, this cannot be easily represented in the `single instruction, multiple data' vector structure that CPUs and GPUs use for accelerating mathematical operations. As such, models are often best described as a series of vector (or matrix) operations that resolve to the size of the vector of observations. Correspondence between parameters and observations are done by `looking up' the correct (as determined by an index per observation) parameter in the parameter array via a \texttt{gather} operation. These \texttt{gather} operations map the parameters to the observations via a set of indices.

This model can be represented (ignoring the construction of the random variables at the county level) in the \texttt{Pyro} probabilistic programming language as
\begin{listing}[H]
\centering
\begin{minipage}{0.9\linewidth}
\begin{minted}[linenos]{python}
with pyro.plate("counties", J):
    a = pyro.sample("a", dist.Normal(gamma_0 + gamma_1 * u, sigma_a))

b = pyro.sample("b", dist.Normal(0., 10.))
sigma_y = pyro.sample("sigma_y", dist.HalfCauchy(5.))

with pyro.plate("data", len(y)):
    county_effects = torch.gather(a, 0, county_idx) 
    mu = county_effects + b * x
    pyro.sample("obs", dist.Normal(mu, sigma_y), obs=y)
\end{minted}
\end{minipage}
\end{listing}

The likelihood is constructed in the final line, within the `data' plate, using observations \texttt{y} of length \texttt{len(y)}. As such, it is within this plate that the county-level effect that corresponds to each observation must be accounted for, which is done via the \texttt{gather} operation. This is otherwise is equivalent to the Python array indexing operation
\begin{listing}[H]
\centering
\begin{minipage}{0.9\linewidth}
\begin{minted}[linenos]{python}
mu = a[county_idx] + b * x
\end{minted}
\end{minipage}
\end{listing}
In either case, the result is an array of length \texttt{len(y)} where the correct county level parameter is present in each element, with repetitions when different measurements have the same county index. If \texttt{county\_idx} were instead a mapping from the counties to homes (\emph{i.e.,} in the reverse direction in Figure~\ref{fig:radon_model_data}B), another completely reasonable way of constructing the indices, nothing would prevent the array shapes from being consistent and the model running successfully. Nevertheless, this would result in nonsense likelihood values, as the homes and counties no longer correspond. As such, even in this simple example, it would be nice to have a mechanism to specify and ensure the \emph{intent} of these otherwise anonymous arrays of indices.

\subsection*{Deeper hierarchies with state-level covariates}

The problem starts to grow when additional levels are introduced to the hierarchy, to describe how county-level effects are driven, in part, by state-level covariates. These might describe data such as state-wide zoning restrictions, or large-scale geological trends. The plate schematic of such a model is given in Figure~\ref{fig:radon_model_data}. In this case, \texttt{gamma\_0} in the previous example is now itself derived from several random variables that depend on the state in question,
\begin{listing}[H]
\centering
\begin{minipage}{0.9\linewidth}
\begin{minted}[linenos]{python}
with pyro.plate("states", S):
    gamma_0 = pyro.sample(
        "gamma_0",
        dist.Normal(eta_0 + eta_1 * v_state, sigma_s)
    )
\end{minted}
\end{minipage}
\end{listing}
Instead of being a scalar, \texttt{gamma\_0} is now a vector of length $S$, and the index vector, \texttt{state\_index}, that maps from states to counties $I_{S \rightarrow C}$ is of length $|C|$, and contains indices into the \texttt{gamma\_0} vector,
\begin{listing}[H]
\centering
\begin{minipage}{0.9\linewidth}
\begin{minted}{python}
a = gamma_0[state_index_a] + gamma_1 * u_county
\end{minted}
\end{minipage}
\end{listing}

From here, the county-level effect is indexed into the array of observations as before. Alternatively, we could formulate this so that the state-level covariates also index directly into the array of observations, via \texttt{state\_index\_obs},
\begin{listing}[H]
\centering
\begin{minipage}{0.9\linewidth}
\begin{minted}{python}
mu = gamma_0[state_index_obs] + gamma_1 * u_county + b * x
\end{minted}
\end{minipage}
\end{listing}

However, here we start to see a problem. If we are given a vector of indices, how are we supposed to know if it maps from the state-level covariates to the county-level covariates, or directly into the observations? Furthermore, and perhaps more dangerously, there is nothing preventing accidental exchange of the \texttt{state\_index\_obs} and \texttt{county\_idx} arrays.

\section{Phantom types}
In programming languages, `phantom' types are parametrised types that do not have a run-time definition. That is, they can be used to enforce strong guarantees about the properties of the parameters of their instances with no run-time overhead. These often specialise more generic fundamental types, permitting algorithmic verification of these guarantees, whilst giving contextual information about the intent of the code to the reader.

For example, it is meaningless to combine physical quantities with different dimensions. However, to the Python interpreter, adding one metre to one kilogram is just as reasonable as adding one metre to two metres, as it has no semantic information other than that these are both equivalently floating point values. By introducing a phantom type \texttt{Quantity} and parameters of this, \texttt{Metre} and \texttt{Kilogram},
\begin{listing}[H]
\centering
\begin{minipage}{0.9\linewidth}
\begin{minted}[linenos]{python}
from typing import Generic, TypeVar

U = TypeVar('U')

class Quantity(Generic[U]):
    def __init__(self, value: float) -> None:
        self.value = value

    def __add__(self, other: 'Quantity[U]') -> 'Quantity[U]':
        return Quantity(self.value + other.value)

class Metre: pass
class Kilogram: pass
\end{minted}
\end{minipage}
\end{listing}

we can then create objects where, for example, only specifically defined operations are allowed (here, addition between objects of the same phantom type). Defining two of these objects,
\begin{listing}[H]
\centering
\begin{minipage}{0.9\linewidth}
\begin{minted}[linenos]{python}
w1 = Quantity[Kilogram](2.0)
d1 = Quantity[Metre](3.0)
\end{minted}
\end{minipage}
\end{listing}
we can then use as normal variables,
\begin{listing}[H]
\centering
\begin{minipage}{0.9\linewidth}
\begin{minted}[linenos]{python}
sum_dists = d1 + d1
sum_weights = w1 + w1
\end{minted}
\end{minipage}
\end{listing}
Importantly, we can verify that these types are compatible by using the static type checking package, \texttt{mypy}:
\begin{listing}[H]
\centering
\begin{minipage}{0.9\linewidth}
\begin{minted}{bash}
> mypy phantom_addition.py
Success: no issues found in 1 source file
\end{minted}
\end{minipage}
\end{listing}

However, if we attempt to add objects of different \texttt{Quantity},
\begin{listing}[H]
\centering
\begin{minipage}{0.9\linewidth}
\begin{minted}{python}
sum_dist_time = w1 + d1
\end{minted}
\end{minipage}
\end{listing}

\texttt{mypy} informs us of a type mismatch, as there is no \texttt{\_\_add\_\_} operation defined for these types:
\begin{listing}[H]
\centering
\begin{minipage}{0.9\linewidth}
\begin{minted}{bash}
> mypy phantom_addition.py
phantom_addition.py:28: error: Unsupported operand types for + 
    ("Quantity[Kilogram]" and "Quantity[Metre]")  [operator]
Found 1 error in 1 file (checked 1 source file)
\end{minted}
\end{minipage}
\end{listing}

We can use this behaviour not only to define specifically allowed operations, but also to define what parametrised types those operations return, given their input types. In this way we can ensure, for example, that when an array is gathered by index, the resulting vector inherits the semantics of the vector from which the indices were derived. 

\section{The typegeist type system}

By introducing a phantom type system to the construction of a hierarchical model, we can ensure correct relationships between the model parameters, avoiding issues such as those identified in the introduction. Moreover, as these types are invisible at run-time, there is no additional overhead. To do this, we first start by a introducing a $\text{Vec}(K)$ type, which represents a vector of model parameters with the identifier $K$, and a $\text{Idx}(K, D)$ type which represents indices from parameter $K$ into dataset $D$. Our aim is ultimately to relate parameter vectors to the entries in the observations vector that they describe, but we may also want to relate these to other parameters in the model, depending on the depth of the hierarchy.

We can introduce operations that convert between types that describe different parameters or datasets. A \texttt{reindex} operation takes indices from $L$ to $K$ within dataset $D$
\begin{equation}
\texttt{reindex}:\ \mathrm{Map}(K,L,D)\times \mathrm{Idx}(L,D)\ \to\ \mathrm{Idx}(K,D),
\end{equation}
and \texttt{lift} transforms parameter vectors of type $K$ into parameter vectors of type $L$
\begin{equation}
\texttt{lift} : \text{Map}(K, L, D) \times \text{Vec}(K) \rightarrow \text{Vec}(L).
\end{equation}
The context for these transformations is the mapping $\text{Map}(K, L, D)$ which describes legal mappings between $K$ and $L$ in $D$. Concretely, we store each map as a 1D array of length $|L|$ with entries in $\{0,\ldots,|K|-1\}$, i.e., a dense `child to parent' array where $\text{array}[l]=k$.

Also provided is a \texttt{gather} operation, to be used when indexing parameters into observations,
\begin{equation}
\texttt{gather} : \text{Idx}(K, D) \times \text{Vec}(K) \rightarrow \text{Array}.
\end{equation}
This is a terminal step, and produces a raw array with no phantom type, as subsequent downstream operations (\emph{e.g.}, optimisation) are not expected to rely on the model semantics.

\begin{figure}
\label{fig:dag}
\centering
\includegraphics[width=0.9\textwidth]{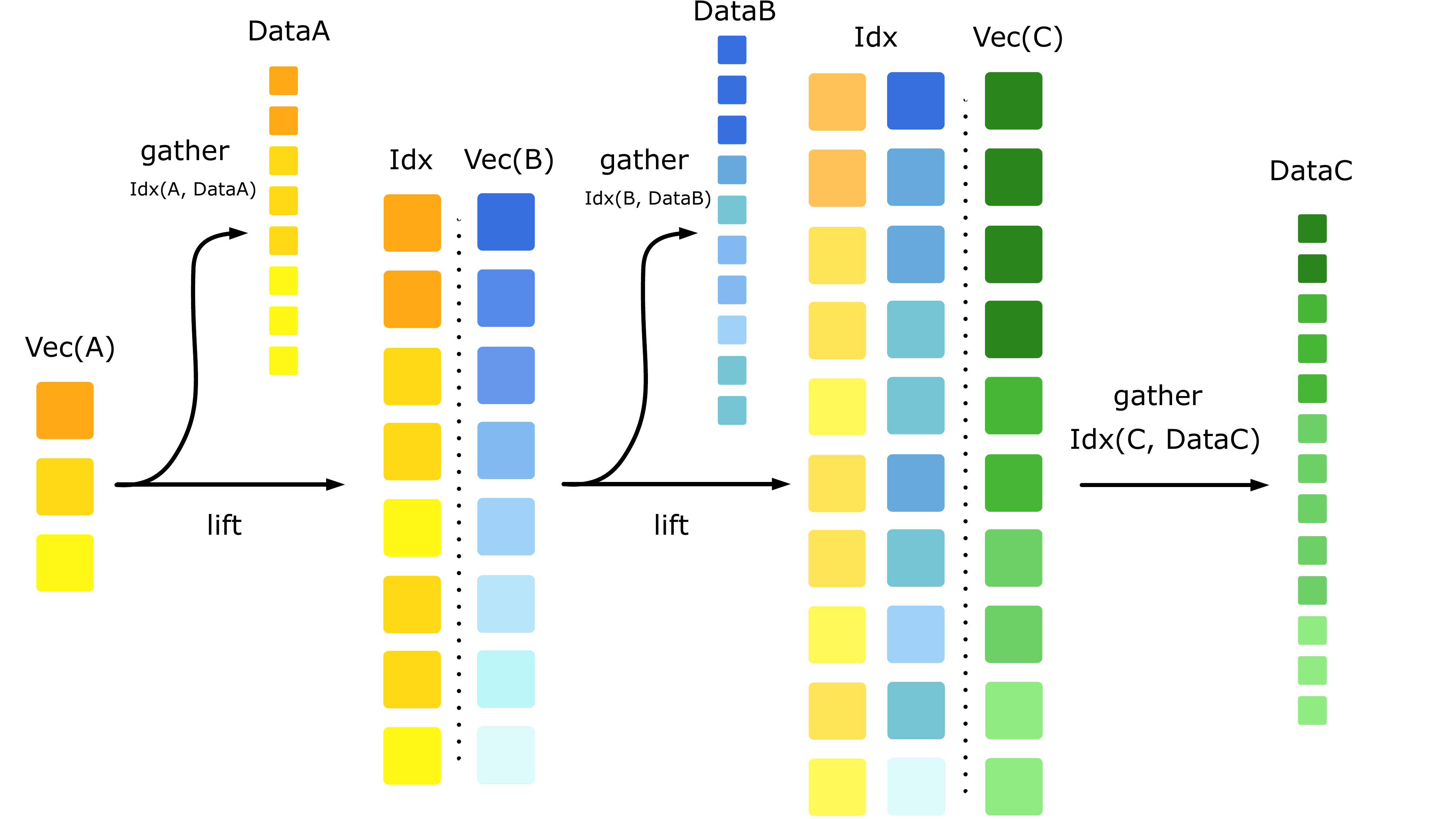}\\
\caption{A directed acyclic graph that describes permitted operations between parameters $A$, $B$, and $C$. Here each has a \texttt{gather} operation to map the random variables to the data that inform them via an index array \texttt{Idx[Param, Data]}. For parameters $A$ and $B$, these are \texttt{lift}ed using indices that map them to parameters lower in the hierarchy, informing their subsequent downstream relationships. In general, multiple parameters can inform each level of the hierarchy, although this is not shown here for brevity.}
\end{figure}

With these definitions, we have developed a system that guarantees consistency of types throughout the parameter reshaping operations. To ensure that these operations perform only intended transformations, we also provide a way to specify those $K$ and $L$ on which $K \rightarrow L$ is valid via $\text{Map}(K, L, D)$. An example schematic of this can be seen in Figure~3, where $\text{Map}(A, B)$ is permitted, however $\text{Map}(A, C)$ would result in a type error even if indices and shapes were consistent.

To enforce these types, we use the \texttt{mypy} static type checker~\cite{mypy}. \texttt{mypy} builds a Python abstract syntax tree, constructs constraints from the type annotations, and solves these using a union-find algorithm which efficiently merges and reconciles equivalent types. If the types of the left and right-hand side of an equation are inconsistent, then \texttt{mypy} emits a type error.

\section{Type guarantees for the radon model}
\label{sec:radon2}

Revisiting the radon example in section~\ref{sec:hm}, we can set up our phantom types for the dataset 
\begin{listing}[H]
\centering
\begin{minipage}{0.9\linewidth}
\begin{minted}{python}
class Data: pass
\end{minted}
\end{minipage}
\end{listing}
and for the parameter annotations
\begin{listing}[H]
\centering
\begin{minipage}{0.9\linewidth}
\begin{minted}{python}
class Home: pass
class County: pass
\end{minted}
\end{minipage}
\end{listing}
as empty classes. We can then define the indices that map uranium measurements in each county to their corresponding observations
\begin{listing}[H]
\centering
\begin{minipage}{0.9\linewidth}
\begin{minted}{python}
county_idx = Idx[County, Data](indices)
\end{minted}
\end{minipage}
\end{listing}
where \texttt{indices} is obtained from the original input dataset. To indicate to \texttt{typegeist} that this index mapping is the one we want to use to go from county-level variables to home-level variables (\emph{i.e.}, at the level of the observations), we register it as a map
\begin{listing}[H]
\centering
\begin{minipage}{0.9\linewidth}
\begin{minted}{python}
register_map(Data, County, Home, county_idx)
\end{minted}
\end{minipage}
\end{listing}
The only operations that are allowed to occur as those that have a registered $\text{Map}$. As such, all other operations on \texttt{typegeist} typed variables (for example, an attempt to map in the reverse direction, from \texttt{Home} to \texttt{County}) will raise an error.

Returning to the \texttt{counties} plate, 
\begin{listing}[H]
\centering
\begin{minipage}{0.9\linewidth}
\begin{minted}{python}
with pyro.plate("counties", J):
    a_raw = pyro.sample("a", dist.Normal(gamma_0 + gamma_1 * u, sigma_a))
\end{minted}
\end{minipage}
\end{listing}
we can define this as a parameter vector with the \texttt{County} annotation
\begin{listing}[H]
\centering
\begin{minipage}{0.9\linewidth}
\begin{minted}{python}
a = Vec[County](a_raw)
\end{minted}
\end{minipage}
\end{listing}

Now we are in a position to use the typed parameter vector to build our likelihood. In this case, the use is fairly trivial, as there is only one operation we wish to perform, the \texttt{gather} to map this to \texttt{Home}, and therefore also the observations,
\begin{listing}[H]
\centering
\begin{minipage}{0.9\linewidth}
\begin{minted}{python}
county_effects = a.gather(county_idx)

with pyro.plate("data", len(y)):
    mu = county_effects + b * x
    pyro.sample("obs", dist.Normal(mu, sigma_y), obs=y)
\end{minted}
\end{minipage}
\end{listing}

In Pyro and NumPyro, we can go one step further and avoid having to explicitly construct the \texttt{Vec} ourselves, by constructing the plate ahead of time with \texttt{AxisPlate},
\begin{listing}[H]
\centering
\begin{minipage}{0.9\linewidth}
\begin{minted}{python}
Counties = AxisPlate("counties", J, axis=County, dataset=Data)
\end{minted}
\end{minipage}
\end{listing}
and using this in the model construction
\begin{listing}[H]
\centering
\begin{minipage}{0.9\linewidth}
\begin{minted}{python}
with Counties:
    a = pyro.sample("a", dist.Normal(gamma_0 + gamma_1 * u, sigma_a))
\end{minted}
\end{minipage}
\end{listing}
where \texttt{a} automatically now has the type \texttt{Vec[County]}.

\subsection*{State-level covariates}

At the state level, there is no additional data, only the model component that represents a hypothesis that county-level observations are linked by a state-level covariate. Nevertheless, there is still a mapping from counties to states, which enables these constraints on the model. This level has the \texttt{State} parameter annotation

\begin{listing}[H]
\centering
\begin{minipage}{0.9\linewidth}
\begin{minted}{python}
class State: pass
\end{minted}
\end{minipage}
\end{listing}

and indices that describe how counties are mapped to states

\begin{listing}[H]
\centering
\begin{minipage}{0.9\linewidth}
\begin{minted}{python}
state_idx = Idx[State, Data](state_idx)
\end{minted}
\end{minipage}
\end{listing}

which we also register a map for

\begin{listing}[H]
\centering
\begin{minipage}{0.9\linewidth}
\begin{minted}{python}
register_map(Data, State, County, state_idx)
\end{minted}
\end{minipage}
\end{listing}

The state-level model is the same, 

\begin{listing}[H]
\centering
\begin{minipage}{0.9\linewidth}
\begin{minted}[linenos]{python}
with pyro.plate("states", S):
    gamma_0_raw = pyro.sample(
        "gamma_0",
        dist.Normal(eta_0 + eta_1 * v_state, sigma_s)
    )
\end{minted}
\end{minipage}
\end{listing}

which we now manually annotate as a \texttt{State} parameter vector

\begin{listing}[H]
\centering
\begin{minipage}{0.9\linewidth}
\begin{minted}{python}
gamma_0 = Vec[State](gamma_0_raw)
\end{minted}
\end{minipage}
\end{listing}

In this example, we don't have explicit mappings from the state-level to the observation level (although we could calculate them with \texttt{Idx} if we wished). This enables a more compact specification of the observation-level likelihood, where only those that have explicit observation indices appear in the expression, and matches what is often the case in most datasets. Therefore in this case, rather than using the \texttt{gather} operation to map the state-level parameters directly to the observation level, we use a \texttt{lift} to map the state-level parameters to the county level,

\begin{listing}[H]
\centering
\begin{minipage}{0.9\linewidth}
\begin{minted}{python}
gamma_0_county : Vec[County] = auto_lift(gamma_0, to=County, dataset=Data)
\end{minted}
\end{minipage}
\end{listing}

We can then use this in the county-level plate

\begin{listing}[H]
\centering
\begin{minipage}{0.9\linewidth}
\begin{minted}[linenos]{python}
with pyro.plate("counties", J):
    a_raw = pyro.sample("a", dist.Normal(gamma_0_county + gamma_1 * u, sigma_a))
a = Vec[County](a_raw)
\end{minted}
\end{minipage}
\end{listing}

where, as before, \texttt{a} is \texttt{gather}ed to the observation shape.

\subsection*{Guarantees}

Even in the simple radon model case, \texttt{typegeist} leads to several guarantees that the model is well formed with respect to the relationships defined by the data. Potential incorrect model specifications are summarised in Table~\ref{tab:typegeist-failure-modes}, along with \texttt{mypy} output when using \texttt{typegeist}.

\begin{landscape}

\begin{table*}[th!]
\centering
\begin{threeparttable}
\caption{Index–related failure modes in the radon model, native vs.\ \emph{typegeist} behaviour.}
\label{tab:typegeist-failure-modes}
\renewcommand{\arraystretch}{1.3}
{\small
\begin{tabularx}{\linewidth}{@{} l >{\raggedright\arraybackslash}p{0.33\textwidth} >{\raggedright\arraybackslash}p{0.33\textwidth} >{\raggedright\arraybackslash}p{0.33\textwidth} @{}}
\toprule
\textsc{Failure mode} & \textsc{Erroneous construction} & \textsc{Native behaviour} & \textsc{Typegeist behaviour (\texttt{mypy} output)} \\
\midrule
Adjacent-level index inversion &
\begin{tabular}[t]{@{}l@{}}
\texttt{a: Vec[County]} \\
\texttt{home\_idx: Idx[Home,Data]}\\
\texttt{a.gather(home\_idx)}
\end{tabular} &
Effects assigned at wrong level; silent bias &
\texttt{error: Argument 1 to "gather" has incompatible type "Idx[Home, Data]"; expected "Idx[County, Data]"} \\

\addlinespace

Unregistered lift &
\begin{tabular}[t]{@{}l@{}}
\texttt{a: Vec[County]}\\
\texttt{auto\_lift(a, to=Unregistered,}\\
\texttt{dataset=Data)}
\end{tabular} &
Wrong semantics &
\texttt{error: Argument "to" to "auto\_lift" has incompatible type "type[Dummy]"; expected "type[Home]"} \\

\addlinespace
Dropped county plate &
Sampling \texttt{a} outside \texttt{plate("counties")} &
Collapses to full pooling across counties &
\texttt{error: Incompatible types in assignment (expression has type "Vec[County]"; variable has type "Vec[Home]")} \\

\addlinespace
Mismatched observation axis &
Combining \texttt{mu\_county} and \texttt{x\_obs} of different lengths &
Likely shape/broadcast error &
\texttt{error: Argument 1 to function has incompatible type "Vec[County]"; expected "Vec[Home]"} \\
\bottomrule
\end{tabularx}
} 
\end{threeparttable}
\end{table*}

\end{landscape}

\section{Summary}

Probabilistic programming frameworks allow the specification and inference of statistical models in mainstream programming languages, enabling expressive descriptions and access to the latest advances in computational techniques. However, whilst these languages offer safeguards against numerous types of programming mistakes, they are largely unaware of the logical consistency of the models they represent. This leads to a significant gap, where the silent mis-specification of model structure or parameter relationships can lead to issues that are hard to debug. The \texttt{typegeist} package for Python enforces parameter relationships within these statistical models, eliminating a common class of modelling errors. Here we have shown how to use this system to guarantee correctness of simple hierarchical models, and how in more complicated models, non-trivial relationships can be easily communicated and verified.

\bibliography{refs}

\end{document}